\documentclass[12pt]{article}
\usepackage[T1]{fontenc} 
\usepackage[margin=1in]{geometry}

\usepackage{amsthm, amsmath, amssymb, amsfonts, mathrsfs, mathtools, amsbsy, latexsym, bbm, bm}

\usepackage{graphicx, epsfig, booktabs, threeparttable, multirow, caption, subcaption, tikz, adjustbox, epstopdf}

\usepackage{setspace, fancyhdr, natbib, url, verbatim, comment, soul}
\usepackage[section]{placeins} 
\usepackage[pagebackref=false]{hyperref}
\hypersetup{
    colorlinks=true,
    citecolor=black,
    filecolor=black,
    linkcolor=black,
    urlcolor=black,
    bookmarksopen=true,
    pdfstartview=FitH
}

\newtheorem{theorem}{Theorem}[section]
\newtheorem{assumption}{Assumption}[section]

\newtheorem{corollary}{Corollary}[section]
\newtheorem{lemma}{Lemma}[section]
\newtheorem{condition}{Condition}[section]

\theoremstyle{definition}

\makeatletter
\def\thm@space@setup{
  \thm@preskip=15pt \thm@postskip=15pt
}
\makeatother

\renewcommand{\qed}{\hfill \mbox{\raggedright \rule{0.08in}{0.08in}}} 
\renewenvironment{proof}[1][\proofname]{{\noindent\sc#1. }}{\qed\vspace{15pt}} 

\long\def\comment#1{}

\newcounter{bean}

\title{Synthetic Control Inference for Staggered Adoption}
\author{Jianfei Cao\thanks{Department of Economics, Northeastern University. E-mail: j.cao@northeastern.edu.} \and Shirley Lu\thanks{Harvard Business School. E-mail: slu@hbs.edu.} \and Hang Wu\thanks{Department of Economics, Northeastern University and School of Management, Fudan University. E-mail: 22110690005@m.fudan.edu.cn.}}

\begin{document}
\maketitle
\onehalfspacing

\begin{abstract}
  {\sc Abstract.} We introduce a synthetic control methodology to study policies with staggered adoption. Many policies, such as the board gender diversity policies, are replicated by other policy setters at different time frames. Our method estimates the dynamic average treatment effects on the treated using variation introduced by the staggered adoption of policies. Our method gives asymptotically unbiased estimators of many interesting quantities and delivers asymptotically valid inference. By using the proposed method and national labor data in Europe, we find evidence that regulation on board gender diversity leads to an increase in full-time employment for female professionals.
  
  
 
\end{abstract}


  \section{Introduction}

    In the policy arena, many reforms are adopted by learning or taking ideas from other policymakers (see \citealp{Dolowitz2000}), which leads to policy being implemented in a staggered way. Examples include the labor market reforms across European countries, and policies sequentially adopted by domestic regions; see, for example, \cite{East2023} and \cite{Eichhorst2017}. This provides opportunities to study the effectiveness of a policy being staggered adopted by multiple policy setters at varying time periods.

In this context, one typical practice is to apply difference-in-differences (DID) or the synthetic control method to panel datasets aggregated at the country or state level, which provide information both before and after policy implementation.

However, the aggregate data used for such studies poses challenges for these methodologies. First, aggregate level information make the assumptions underlying staggered DID methods less credible. Recent econometric literature has extended DID methods to settings with dynamic treatment effects in order to address the issue of negative weights of the treatment effects; see, e.g.\ \citet{DeChaisemartin2020,Goodman-Bacon2021,Sun2021,Callaway2021}. Crucially, these methods assumes parallel trends across units, which may fail to hold in aggregated-level data. For example, we may not expect female employment in Norway to have the same trend as that in the United Kingdom.

The second commonly used method, synthetic control, mitigates the problems above, and are often used in comparative case studies with a moderate number of  aggregated units; see \cite{Abadie2010}. However, there is an issue when synthetic control is used in a staggered adoption setting: the number of units in the donor pool is greatly reduced. Often, the best candidates for synthetic control are those units that were treated at a different time period, yet current synthetic control methods only use units that are never-treated as synthetic control candidates. In settings where  majority of the units are treated at some point in time, synthetic control method may fail to construct a “good” synthetic control unit.

In this paper we propose a new synthetic control method to study heterogeneous treatment effect of policies with staggered adoption, along with an inference procedure based on Andrews’ instability test. Our approach does not rely on parallel trends across units as in DID methods, and it extends the synthetic control frame work by using all important units, including not-yet treated, to form synthetic control. The idea is that the counter-factual outcome in treated periods should behave similarly to the untreated periods. To achieve that, we estimate a model for each unit using all other units subject to the synthetic control constraint, and then simultaneously estimate all unit × time treatment effects. We formalize the underlying assumptions and establish the asymptotic properties for the proposed methods. 

While we mainly focus on the dynamic effects of policies by looking at the average treatment effects on the treated for some periods after treatment (event-time ATT),  our framework can be readily extended to other parameters of interest. That is, beyond tracking how effects evolve over time, the proposed method allows comparison of effects across groups (group-time ATT) or policies.
 
We apply the staggered synthetic control method to study the effects of board gender diversity policies introduced in a staggered time frame across 14 countries in the EU. Two challenges motivate the use of staggered synthetic control in this setting. First, we are interested in the long run impact of the policies on employment outcomes. Since these outcome variables are available at the aggregated country level, the moderate number of observations may not fulfill the large sample assumption used in generalized diff-in-diff asymptotics. 
Second, there are only few European countries with no policy yet, such as Cyprus and Malta, which tend to be smaller and less comparable to the treated European countries, such as France and Belgium. As such, using traditional synthetic control, we have limited control candidates.  

Using our proposed methodology, we show that corporate board female ratio increased significantly after the policy announcements. This increase is realized gradually from an average of 6\% increase in the first year after the policy announcement, to 11\% increase after four years. Moreover, in the long term, policies implemented via quotas lead to greater effect than those implemented via disclosure mandates.

Our paper is related to two streams of literature. First, there is a rising literature on estimating treatment effects for policies with staggered adoption. Several papers extend DID method to accommodate this setting; see, e.g. \citet{Sun2021,Borusyak2024,Callaway2021,DeChaisemartin2020,Goodman-Bacon2021}. Among them, the method proposed by \citet{Callaway2021} is the most similar to ours in spirit. They estimate the block-level treatment effects and then construct estimators for interesting parameters, allowing for great flexibility. However, all of the aforementioned works rely on some strong notion of common trend, which is likely to fail in many cases. 

Second, some literature  works on the multivariate synthetic control method; see, e.g.  \citet{Firpo2018,Kreif2016,Robbins2017,Xu2017}. This literature applies the synthetic control method to cases where more than one unit is treated by the policy. Moreover, several studies propose inference procedure for this framework. \citet{Abadie2021} and \citet{Arkhangelsky2021} address cases with simultaneous adoption of treatment. 
\citet{Ben-Michael2022} introduce a bootstrap procedure for event-time ATT estimator, though without theoretical results. \citet{Cattaneo2025} propose non-asymptotic prediction intervals. However, these approaches typically throw away not-yet-treated units when forming weights, which can lead to efficiency loss. Exceptions include \citet{Donohue2019} and \citet{Powell2022}, who incorporate not-yet-treated units; yet \citet{Donohue2019} dooes not provide formal justification, and the Wald-based inference by \citet{Powell2022} relies on the implicit assumption of long post-treatment periods. Our contribution is twofold: (1) we adapt the synthetic control methodology to better study dynamic effects by allowing not-yet-treated units to be part of the synthetic control; (2) we present an inference procedure based on Andrews’ end-of-sample instability test that is suitable for small numbers of units and short post-treatment periods, delivering valid inference for several parameters of interest.


The rest of the article is organized as follows. Section 2 describes the main ideas behind the staggered synthetic control approach to policy interventions implemented in a staggered time frame. In Section 3 we apply the proposed staggered synthetic control methods to estimate the effect of board gender diversity policies in the EU. 
Section 4 concludes. 
All proofs are given in Appendix A.
Appendix B provides two examples of primitive assumptions that justify our assumptions in the main text. 

\section{Synthetic Control Methods for Staggered Adoption of Policies}

In this section, we propose a framework that enables estimation of various parameters of interest including event-time ATT (defined in Section \ref{subsection_rubin_model}). Our inference procedure is based on Andrews' end-of-sample instability test. We give formal assumptions under which the proposed procedure has asymptotically correct size.

\subsection{A Rubin model}
\label{subsection_rubin_model}

Consider a panel of $N\times (T+S)$ observed for outcome variable $y_{i,t}$ and treatment status $d_{i,t}$, where $N$ is the number of units, $T$ is the number of periods before any unit is treated, and $S$ is the number of periods when at least one unit is treated. Throughout, assume $T\rightarrow \infty$ and  $S$ and $N$ are fixed. That is, we assume the pre-treatment data is rich enough to be informative about the relationship among units. 

Consider a version of Rubin's potential outcome model:
\begin{equation*}
y_{i,t}=\begin{cases}
y_{i,t}(1),\text{ if } d_{i,t}=1,\\
y_{i,t}(0),\text{ otherwise.}
\end{cases}
\end{equation*}
For $t=1,\dots,T$, no units are treated, i.e., $d_{i,t}=0$ if $t\le T$. Assume that the unit is always treated once it has been treated, i.e., $d_{i,t}\le d_{i,s}$ if $t\le s$. 
Note this framework assumes the Stable Unit Treatment Value Assumption (SUTVA). That is, the outcome of unit $i$ at time $t$ is only a function of its own treatment status at this time period, and timing of the treatment and treatment status of any other unit at any periods does not impact the outcome.

The individual treatment effects are defined by 
$$\tau_{i,t}=y_{i,t}(1)-y_{i,t}(0)$$
for $(i,t)$ with $d_{i,t}=1$. 
We allow for full flexibility on the form of treatment effects and do not impose any parametric restriction. 
Let $\tau_{i,t}=0$ for $d_{i,t}=0$ for notation simplicity. 
Let $\tau\in \mathbb{R}^K$ be the vectorization of $\tau_{i,t}$'s such that unit $i$ has been treated at time $t$, i.e., $\tau=(\tau_{i,t})_{(i,t)\in D}$ where $D=\{(i,t):d_{i,t}=1 \}$. Our parameter of interest is $\gamma=L\tau$ for some linear transformation $L$. Examples including event-time ATT and difference of various types of treatment assignments are given as follows.\footnote{For more examples on interesting parameters, see \citet{Callaway2021}, who look at the difference-in-differences estimator with staggered adoption. }

\subsubsection{Event-time ATT}

Researchers are often interested in estimating the average treatment effects on the treated $s$ periods after being treated (ATT of event time $s$). Define event time 
$$e_{i,t}=\sum_{r\le t}\mathbbm 1\{d_{i,r}=1 \}$$
and number of observations with event time $s$
$$n_s=\sum_{i,t}\mathbbm{1}\{e_{i,t}=s \}.$$
Then the ATT of event time $s$ is 
$$\mathrm{ATT}_s^e=\frac{1}{n_s}\sum_{i,t} \tau_{i,t}\mathbbm 1 \{ e_{i,t} = s \}=l_s'\tau$$
for $l_s=n_s^{-1}(\mathbbm 1\{ r_{i,t} = s \})_{(i,t)\in D}$. One may want to test the hypothesis $H_0:\mathrm{ATT}_s^e=l_s'\tau=0$.

\subsubsection{Comparison among policies}

It is often important to evaluate the difference in treatment effects induced by various types of policies. For example, one may want to know how the dynamic treatment effects of \emph{Quota} differ from those of \emph{Disclosure} in terms of female board members. In our set-up, suppose there are two types of policy implemented. For $i=1,2$, the event-time ATT at time $s$ of the $i$-th policy can be written as $\text{ATT}_{i,s}^e=l_{i,s}'\tau$ for some $l_{i,s}$ properly defined. Then, one may want to estimate $(\mathrm{ATT}_{1,s}^e,\mathrm{ATT}_{2,s}^e)'=[l_{1,s}, l_{2,s}]'\tau$, and be interested in testing the hypothesis $H_0:\text{ATT}_{1,s}^e-\text{ATT}_{2,s}^e=C\tau=0$, where $C=l_{1,s}'-l_{2,s}'$.

\subsection{An invertibility assumption}



Since we assume the number of units and the number of post-treatment time periods are small relative to the number of pre-treatment time periods (large $T$, small $S$ and $N$), the individual treatment effects are not point-identified from the data. In order to learn useful information about the treatment effect, we discuss a key assumption in this section. 

We first define the individual synthetic control weights and their limits. Namely, let 
\begin{equation}
\label{optimization}
\begin{bmatrix}
\hat{a}_i\\\hat{b}_{i}
\end{bmatrix}=\underset{(a,b)\in W_i}{\arg\min}\sum_{t=1}^T(y_{i,t}-a-Y_tb')^2,
\end{equation} 
where $W_i=\{\beta=(\beta_0,\beta_1,\dots,\beta_N)'\in \mathbb{R}\times\mathbb{R}^{N}_+: \beta_i=0, \sum_{j=1}^N\beta_j=1 \}$. Then, let
\begin{equation*}
a_i=\text{plim}\ \hat{a}_i,\ b_i=\text{plim}\ \hat{b}_i,
\end{equation*} 
and we only consider cases where they are well-defined (see Appendix~\ref{low_level_section} for primitive assumptions under which $\hat{a}_i$ and $\hat{b}_i$ converge).

For each $i$ and $t$, define the ``prediction error'' by 
\begin{equation}
\label{linear combination form}
u_{i,t}=y_{i,t}(0)-(a_i+Y_t(0)'b_i).
\end{equation}
Note that the $i$-th entry of $b_i$ is zero. Define $a=(a_1,\dots,a_N)'$, $B=(b_1,\dots,b_N)'$, and $M=(I-B)'(I-B)$. Also define their sample analog $\widehat{a}=(\widehat{a}_1,\dots,\widehat{a}_N)'$, $\widehat{B}=(\widehat{b}_1,\dots,\widehat{b}_N)'$, and $\widehat{M}=(I-\widehat{B})'(I-\widehat{B})$. For each $s=1,\dots,S$, further define linear transformation $A_s\in \mathbb{R}^{N\times K}$ such that $\tau_s=A_s\tau$, where $\tau_s=(\tau_{i,t})_{t=T+s}$ is the $N$-dimensional ``effect vector'' at time $T+s$. We introduce the following invertibility assumption:

\begin{assumption}
	\label{identification_assumption}
	(Invertibility) $\sum_{s=1}^SA_s'MA_s$ is invertible. 
\end{assumption}

Assumption \ref{identification_assumption} excludes cases where all units are treated at a certain time period $t$. This assumption identifies the distribution of some population quantity that is centered at the true parameter, which facilitates the (asymptotic) unbiased estimation (showed in Section \ref{subsection_estimation}). Note that this assumption is testable in principle.

\subsection{Estimation of parameters of interest}
\label{subsection_estimation}

As discussed in the previous section, the treatment effects are not identified, so consistent estimation is impossible. In this section, we propose an asymptotically unbiased estimator of the parameter of interest. 

Stacking equation \eqref{linear combination form} for all $i$'s gives 
\begin{equation*}
u_t=Y_t(0)-(a+BY_t(0)),
\end{equation*}
where  and $u_t=(u_{1,t},\dots,u_{N,t})'$. After the first unit is treated and for some $s=1,\dots,S$, this becomes 
\begin{equation}
\label{error equation}
u_{T+s}=(I-B)(Y_{T+s}-\tau_s)-a.
\end{equation} 

Let $\Vert \cdot \Vert$ be the Euclidean norm and $\Vert \cdot \Vert_F$ be the Frobenius norm.

\begin{assumption}
	\label{unbiased_assumption}
	(a) $\{u_{t}\}_{t\ge 1}$ is strictly stationary and has mean zero. 
	
	(b) $\Vert \widehat{a}-a\Vert=o_{p}(1)$, $\Vert \widehat{B}-B\Vert_F=o_{p}(1)$, and $\Vert (\widehat{B}-B)Y_{T+s}(0)\Vert=o_{p}(1)$ for each $s$.
\end{assumption}

Assumption 2 implies there is an underlying stationary sequence we can leverage. We will use the stationarity of this sequence to show asymptotic unbiasedness. Part (b) requires the pre-treatment data identifies the prediction model among all units. We show that this assumption holds under stationary or co-integrated common factors when $\{y_{i,t}(0)\}$ follows a factor structure. See Appendix \ref{low_level_section} for details.

The estimator for $\tau$ is given by
\begin{eqnarray}
\label{estimator_formula}
\widehat{\tau} &=& \underset{g\in \mathbb{R}^K}{\arg\min}\sum_{s=1}^S \Vert (I-\widehat{B})(Y_{T+s}-A_s g)-\widehat{a}\Vert_2^2 \notag \\
&=& \Bigg( \sum_{s=1}^S A_s' \widehat{M} A_s \Bigg)^{-1} \Bigg( \sum_{s=1}^S A_s'(I-\widehat{B})'((I-\widehat{B})Y_{T+s}-\widehat{a}) \Bigg). 
\end{eqnarray}
Then the estimator for $\gamma=L\tau$ is  
$$\widehat{\gamma}=L\widehat{\tau}.$$

\begin{theorem} 
	\label{estimation_theorem}
	Suppose Assumption \ref{identification_assumption} and \ref{unbiased_assumption} hold. Then, as $T\rightarrow \infty$,
	$$\widehat{\gamma}-(\gamma +LV_T)\stackrel{p}{\rightarrow} 0,$$
	where $$V_T=\Bigg( \sum_{s=1}^S A_s' {M} A_s \Bigg)^{-1} \Bigg( \sum_{s=1}^S A_s'(I-B)'u_{T+s}\Bigg) $$
	and $E[V_T]=0$.
\end{theorem}

That is, $\widehat{\gamma}$ is an asymptotically unbiased estimator for $\gamma$.\footnote{The variance of this estimator can be lowered by estimating a covariance matrix and using it as a weighting matrix. See \citet{Cao2019} for details.} 

In the case of event-time ATT, the estimator is 
$$\widehat{\mathrm{ATT}}_s^e=\frac{1}{n_s}\sum_{i,t} \widehat{\tau}_{i,t}\mathbbm 1 \{ R_{i,t} = s \}=l_s'\widehat{\tau}$$

\begin{corollary}
	\label{inference_att}
	Under Assumption \ref{identification_assumption} and \ref{unbiased_assumption},
	$$\widehat{\mathrm{ATT}}_s^e-(\mathrm{ATT}_s^e +l_s'V_T)\stackrel{p}{\rightarrow} 0$$
	with $E[l_s'V_T]=0$, as $T\rightarrow \infty$. 
\end{corollary}

\subsection{Inference methods}
\label{inference}

We consider a set of linear restrictions on the parameter $\tau$:
$$H_0:C\tau=d,$$
where $d$ is a column vector and $C$ is some fixed matrix. This includes a set of linear restrictions on the parameter of interest $\gamma$ as a special case such as $H_0:{C}\gamma={C}L\tau={d}$ for some fixed $C$. For example, we are often interested in testing whether some policy is effective in terms of event-time ATT:
$$H_0:\mathrm{ATT}_s^e=l_s'\tau=0,$$
which corresponds to the case with $C = 1$  and $L = l_s'$.
Another example is to test whether two policies have significantly different event-time ATT: 
$$H_0:\mathrm{ATT}_{1,s}^e-\mathrm{ATT}_{2,s}^e=(l_{1,s}'-l_{2,s}')\tau=0,$$
which corresponds to ${C} = [1, -1]$  and $L = [l_{1,s}, l_{2,s}]'$.


We propose a test that is based on Andrews' test as in \citet{Andrews2003}. Define the test statistic 
$$\widehat{P}=(C\widehat{\tau}-d)'(C\widehat{\tau}-d).$$
The statistic is expected to be large when the null hypothesis does not hold. 

To form the critical value, first define $x_t=(1,y_{1,t},\dots,y_{N,t})'$. For some $\theta\in \mathbb{R}^{N\times (N+1)}$, let 
$$V_{t}(\theta)=\Bigg( \sum_{s=1}^SA_s'{M} A_s \Bigg)^{-1} \Bigg( \sum_{s=1}^{S} A_s'(I-{B})'(Y_{t+s}-\theta x_{t+s}) \Bigg)  $$
and its sample analog
$$\widehat{V}_{t}(\theta)=\Bigg( \sum_{s=1}^SA_s'\widehat{M} A_s \Bigg)^{-1} \Bigg( \sum_{s=1}^{S} A_s'(I-\widehat{B})'(Y_{t+s}-\theta x_{t+s}) \Bigg).$$
Note that under $H_0$, the limiting distribution of $\widehat{P}$ can be approximated by $V_t(\theta_0)'C'CV_t(\theta_0)$ with $\theta_0=(a,B)$. For a sequence of estimators $\{\widehat{\theta}^{(t)} \}_{t=1}^T$, define 
$$\widehat{P}_t=\widehat{V}_t(\widehat{\theta}^{(t)})'C'C\widehat{V}_t(\widehat{\theta}^{(t)}).$$
In practice, one can simply let $\widehat{\theta}^{(t)}=(\widehat{a},\widehat{B}) $ for each $t$. Another choice is to use the leave-$S/2$-out estimator; see \citet{Andrews2003}. 
For some $\theta\in \mathbb{R}^{N\times (N+1)}$, define
$${P}_t(\theta)={V}_t(\theta)'C'C{V}_t(\theta)$$ 
and its sample analog 
$$\widehat{P}_t(\theta)=\widehat{V}_t(\theta)'C'C\widehat{V}_t(\theta).$$
Note that $\widehat{P}_t=\widehat{P}_t(\widehat{\theta}^{(t)})$. The empirical distribution of $\widehat{P}_t$ is then
$$\widehat{F}(x)=\frac{1}{T-S}\sum_{t=1}^{T-S}\mathbbm 1\{ \widehat{P}_t\le x \},$$
and the corresponding $(1-\alpha)$-quantile is 
$$\widehat{q}_{1-\alpha}=\inf\{x\in \mathbb{R}:\widehat{F}(x)\ge 1-\alpha \}.$$

For some significance level $\alpha$, we reject the null hypothesis if $\widehat{P}$ lies outside the $(1-\alpha)$-quantile of the empirical distribution formed by $\{ \widehat{P}_t\}_{t=1}^{T-S}$, i.e. reject $H_0$ if $\widehat{P}>\widehat{q}_{1-\alpha}$. The confidence region can be constructed by inverting the test. 

We impose an additional assumption to ensure valid inference.
\begin{assumption}
	\label{inference_assumption}
	(a) $\{u_t\}_{t\ge 1}$ is ergodic and has finite second moment. 
	
	(b) There exists a non-random sequence of positive definite matrices $\{D_T \}_{T\ge 1}$ such that $\max_{t\le T+S}\Vert D_T^{-1}x_t\Vert=O_{p}(1)$.
	
	(c) $\Vert (\widehat{\theta}-\theta_0)D_T\Vert_F=o_{p}(1)$, and $\max_{t=1,\dots,T} \Vert (\widehat{\theta}^{(t)}-\theta_0)D_T\Vert_F=o_{p}(1)$, where $\Vert \cdot \Vert_F$ is the Frobenius norm. 
	
	(d) The distribution function of $P_1(\theta_0)$ is continuous and increasing at its $(1-\alpha)$-quantile.
\end{assumption}

Assumption \ref{unbiased_assumption} and \ref{inference_assumption} are similar in spirit to those given by \citet{Chernozhukov2025}. It is worth noting that those assumptions do not preclude methodologies other than the standard synthetic control method. We focus on the synthetic control method because it has good performance when only moderate size datasets are available, which is common in comparative case studies.  When appropriate, one can easily extend our framework to incorporate other estimators. 
Besides, we show by Lemma \ref{lemma_low_level_assumptions} in Appendix \ref{low_level_section} that Part (a)-(c) in Assumption \ref{inference_assumption} are satisfied by either stationary or co-integrated common factors. 

\begin{theorem}
	\label{inference_theorem}
Suppose Assumption \ref{identification_assumption}, \ref{unbiased_assumption}, and \ref{inference_assumption} hold. Then, under $H_0$,
$$\Pr(\widehat{P}>\widehat{q}_{1-\alpha})\rightarrow \alpha,$$ 
as $T\rightarrow \infty$. 
\end{theorem}

That is, even if we are not able to point-identify the parameter of interest, we can derive the asymptotic distribution of the estimator, based on which we can conduct valid inference. Valid confidence intervals can be constructed by inverting the test.

\subsection{Implementation}

In summary, to get the consistent estimator of treatment effect and conducting hypothesis test, our procedure includes the following steps:

\setcounter{bean}{0}
       \begin{center}
        \begin{list}
         {\textsc{Step} \arabic{bean}.}{\usecounter{bean}}
         \item Define parameters to be estimated: synthetic control specification, vectorized treatment effects $\tau$ and $\tau_s$ for each post-treatment period $s$, and parameter of interest $\gamma$.
         Define linear transformation $A_s$ and $C$, with $\tau_s = A_s\tau$ and $\gamma = C\tau$.
         \item Estimate synthetic control weights by solving \eqref{optimization}.
         Calculate the estimator of vectorized treatment effect  $\widehat{\tau}$, by plugging synthetic control weights into \eqref{estimator_formula} and solving the minimization.
         Calculate the estimator of the parameter of interest as $\widehat{\gamma} = C\widehat{\tau}$.
         \item Specify the null hypothesis $H_0$ and the associated test statistics $\widehat{P}$.
         Construct empirical distribution of test statistics $\widehat{F}(x)$ and its $(1-\alpha)$ quantile  $\widehat{q}_{1-\alpha}$.
         Conduct the hypothesis test or construct confidence intervals using the observed statistic and the empirical quantile.
        \end{list}
       \end{center}

\section{Estimating the Effects of Board Gender Diversity Policies}
In this section, we apply the staggered synthetic controls method to a widely adopted policy intervention that has generated considerable debates: gender equality regulation on corporate boards. Proponents argue that a systematic change is needed to address the glass ceiling (see \citealp {EU2012}, hereafter the EU Impact Assessment). Opponents claim that these policies may only benefit the few female directors, as known as ``the golden skirt'', given the supply constraint of female directors (see \citealp{Huse2011,Seierstad2011}), and prior research finds a decrease in firm performance subsequent to these policies (see \citealp{Ahern2012}). One way to reconcile the two is to consider the dynamic effects of these policies: do we observe negative short term consequence on firm performance, in exchange for a better long term societal gender equality. Therefore, we study the implication of the policy both in the longer term and on the wider society.  

Despite the importance to study societal impacts of these policies, most existing papers investigate the policies' impact on firm performance. One reason is that wider societal effects may take time to be realized, making it difficult to measure. Another reason is these outcome variables, such as labor employment, are usually observed at the aggregate level, making it difficult to study in one country. To mitigate these concerns, in this paper, we exploit the staggered adoption of board gender policies in European countries, and apply the proposed synthetic control method for staggered adoptions to examine the magnitudes of the dynamic effects of these policies.

\subsection{Background}
Gender equality has made great progress in the past century, but while the gender gap on higher education and entry-level employment has reduced, the gap at the higher business decision making roles remains large; see \citet{MarianneBertrandClaudiaGoldin2011}. The EU impact assessment in 2012 claims that women account for 60\% of new university graduates who enter the work force, 35\% of the European parliament members, but merely 13.7\% of corporate board seats in large listed companies.

To address the low female representation on corporate boards, Norway first proposed a 40\% quota in 2002. Between 2007 and 2017, 13 European countries adopted similar policies in a staggered manner. These policies are either a board gender quota, or a disclosure mandate requiring firms to discuss board gender diversity plans and progress. Outside the EU, California signed a board gender quota bill in 2018 (Los Angeles Times; see \citealp{McGreevy2018}), and several other US states have considered similar policies.

Many papers criticize the immediate damage brought by a gender quota. From a firm perspective, \citet{Ahern2012} finds that the stock price dropped after Norway’s quota proposal, and that the number of public firms declined, a pattern suggesting avoidance behavior. However, \citet{Eckbo2022} finds no such effect. \citet{Lu2019} finds that, due to the supply constraint, firms subject to the quota hire more foreign or less experienced female directors.

However, short-run evidence of policy distortions alone cannot speak to the desirability of the policies. It is important to study the intended benefits of a board gender policy, especially in the long run. \citet{Bertrand2018} study the effects seven years after Norway's quota policy and show that quotas benefited female directors but did not extend to other positions over time, which was also a goal of the policy. 

Our goal is to study the long term societal effects of board gender equality policies, similar to \citet{Bertrand2018}, and expand to settings involving all European countries that announced such a policy. While many papers focus on specific countries, very few study across countries.

There are three channels a board gender policy can bring long term societal benefits. First, firms may invest in a pipeline for female leadership, such as mentorship programs. Second, the presence of female directors may encourage younger female employees to invest in career development by showing paths to top positions. Third, female leaders can empower junior employees through mentoring and creating a more supportive work environment; see \citet{Dezso2012}. For example, Sheryl Sandberg demanded pregnancy parking spots at Facebook, while such female-oriented policies may never be heard if demanded by non-leadership women. All three channels suggest that in the long term, there may be increased female participation in the workforce and a higher likelihood of women pursuing and attaining leadership positions. 

In the long term, many short term consequences may diminish if the supply of qualified female leaders increases. However, some critiques remain and could limit the aforementioned benefits. For instance, boards may represent only a small segment of society, reducing broader impact, and board members selected based on token rather than merit may be less effective in driving company change, as shown by \citet{Leszczynska2018}. 

Therefore, it is an empirical question whether a board gender policy can lead to long term gender equality benefits in the labor market.

\subsection{Data and Sample}
In this paper, we use two sources of publicly available, country level data in the EU. For the analysis on corporate board's female representation, we use data from the European Institute for Gender Equality (``EIGE''), which is an autonomous body of the European Union dedicated to the promotion of gender equality. We use the percentage of female corporate board members among the largest listed companies in each country, and the data is available annually from 2003 to 2019.

To study how the board gender policies affect female's work decisions, we use labor outcome variables from the EU Labor Force Survey (``LFS''), which is the largest household survey on annual and quarterly employment status covering 35 European countries. Aggregated country level data is available on Eurostat, the statistical office of the EU. While there are a wide range of data related to gender diversity in the LFS database, data quality varies. After examining the quality and completeness of the data, we use full time employment to proxy for changes in the extensive margin, and use weekly work hours to proxy for changes in the intensive margin. These data are mostly available quarterly from 2003Q1 to 2018Q4. 

\subsection{Results}

\subsubsection{Board gender ratio}

\begin{figure}[ht]
\centering
{\includegraphics[width=\textwidth,trim={0cm 2cm 0cm 3cm}]{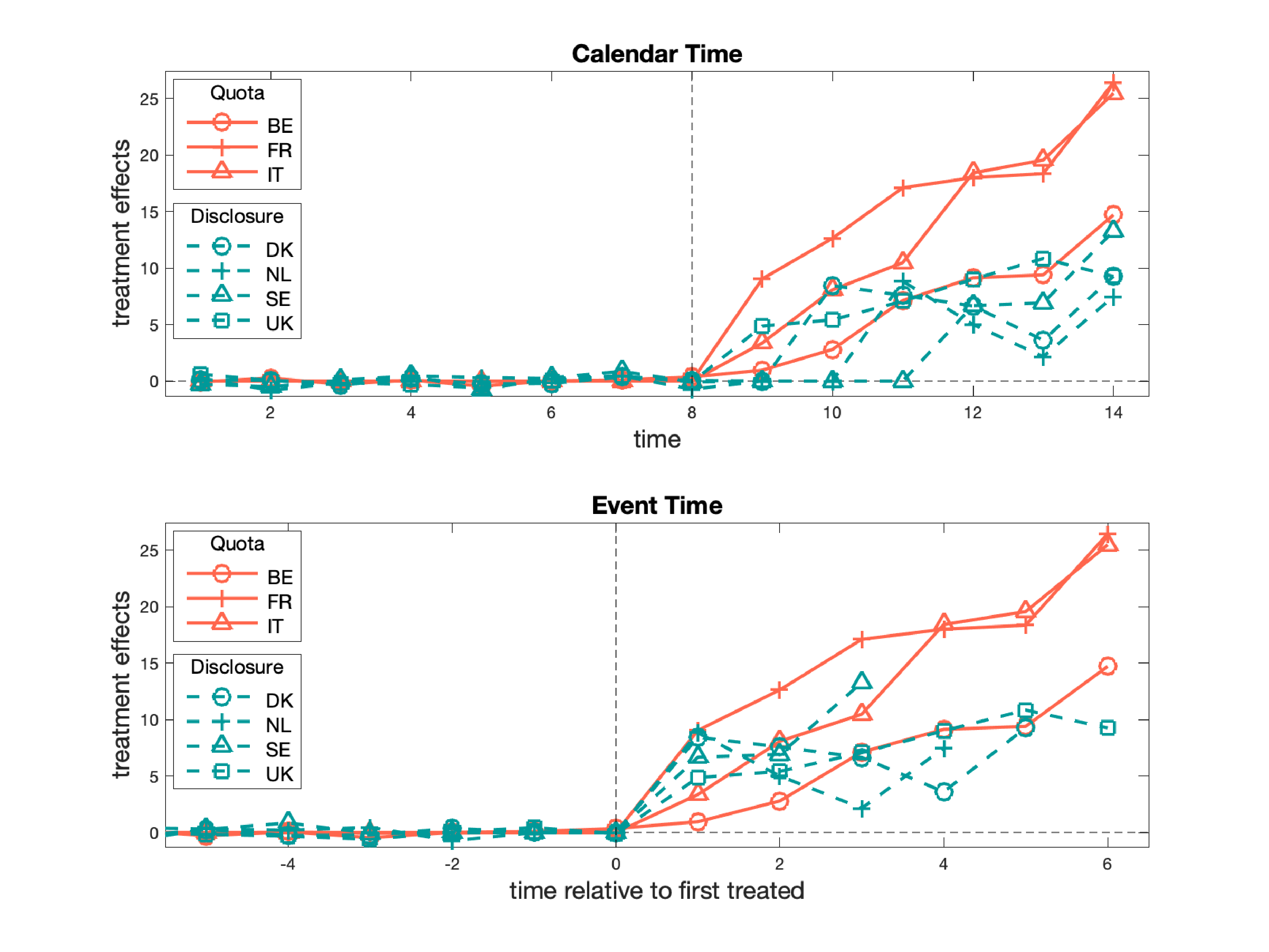} }

\caption{\label{fig:female_ratio_compare}Individual dynamic treatment effects on the percentage of female on corporate boards for quota and disclosure policies separately. Event time is relative to the announcement of board gender diversity policies. All values are shown in percent.}
\end{figure}

We compare the two types of board gender diversity policies: quota vs disclosure. Figure \ref{fig:female_ratio_compare} shows that the effect from a disclosure policy is larger in the first year, but plateaus after. In contract, quota effects increase over the years to over 20\% in the sixth year. One reason for the difference is that quotas include binding targets in a future year, where penalty is involved for failing to reach the targets. 

The treatment effects are relative to the synthetic control for each unit. As an example, Table \ref{table_weights} lists the synthetic control weighting for France. We can see that France's synthetic control is a combination of Belgium, Bulgaria, Croatia, Denmark, Latvia, Lithuania, Portugal, and Sweden, among which Belgium has the highest weight of 49\%. Although we should not over-interpret the estimated weights, the weights for France look reasonable, given that Belgium neighbors France and the two countries share many similarities in terms of the legal system. 

In constructing the sample for this test, we are left with 27 countries. To allow a sufficient pre-treatment time period to estimate the synthetic control weightings, we excluded four countries that announced a policy before 2010. Given the small number of countries, standard large sample asymptotics with respect to units does not apply. Specifically, the popular generalized difference-in-difference method is not ideal. It also suffers from problems such as assuming strong homogeneity in order to produce interpretable results, and requiring parallel trend. 

\begin{table}
\caption{\label{table_weights}Synthetic Controls Weights for France.}
\begin{center}
\begin{tabular*}{0.65\textwidth}{@{}lclc@{}}
\hline\hline
Country        & Weight    & Country        & Weight    \\
\hline
Austria        & 0            &  Latvia & 	0.1340         \\
Belgium        & 0.4931      &Lithuania & 	0.0246        \\
Bulgaria       & 0.0198      &Luxembourg & 	0              \\
Croatia        & 0.1074      & Malta & 	0            \\
Cyprus         & 	0        & Netherlands    & 0    \\
Czech Republic &	0        &Poland         & 0         \\    
Denmark        & 0.0305      &Portugal       & 0.0001        \\
Estonia	       & 0           &Romania       & 0              \\
Germany        & 0           &Slovakia       & 0            \\
Greece         & 0           &Slovenia          & 0         \\
Hungary        & 0           &Sweden         & 0.1904        \\
Ireland        & 0           &United Kingdom & 0       \\
Italy          & 0  \\
\hline\hline
\end{tabular*}
\end{center}
\footnotesize
\renewcommand{\baselineskip}{11pt}
\textbf{Note:} This table lists the country weightings for the construction of France's synthetic control in the analysis on board female ratios.
\end{table}

\subsubsection{Employment outcomes}

Next, we apply the same methodology on labor outcomes to study if board gender diversity policies have wider societal impacts. In particular, we are interested in finding whether these policies encourage females to invest in their career. Since these encouragements likely apply to female who may break the glass ceiling, 
we focus on the professionals occupation as the group most directly affected by these policies. These include professionals in a variety of professions, such as business and administration, IT, science, and legal professionals.\footnote{These categories are based on the International Standard Classification of Occupations (ISCO), as provided in Eurostat.}

\begin{figure}[ht]
\centering
{\includegraphics[width=\textwidth,trim={0cm 7cm 0 6cm}]{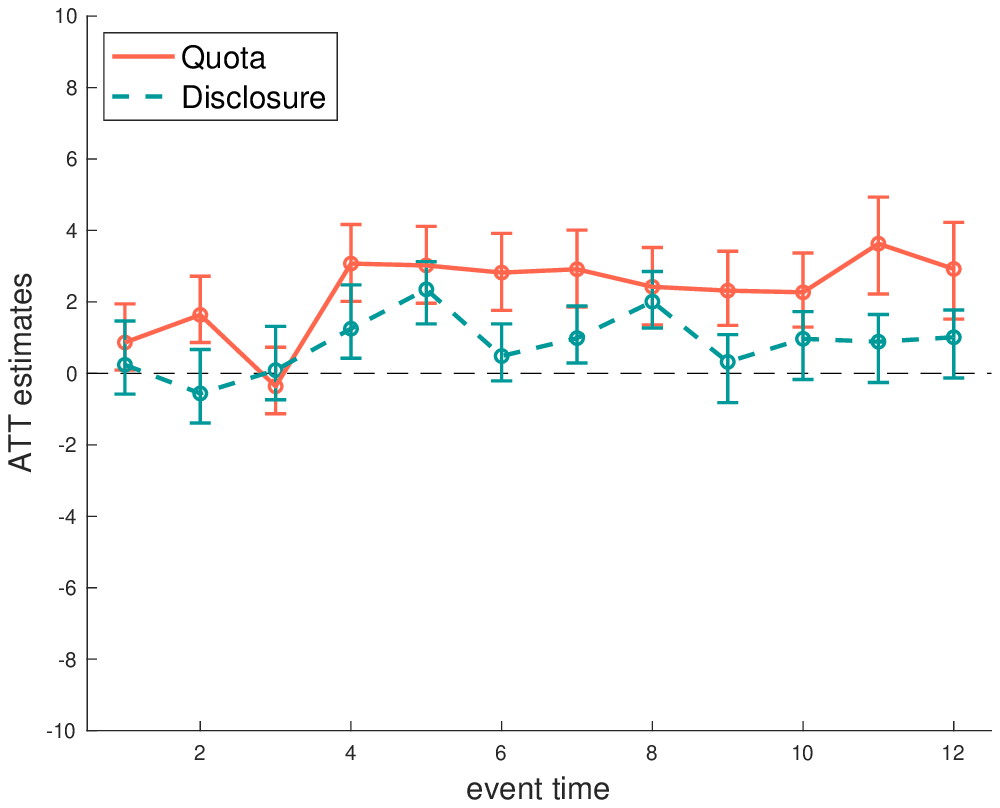} }
\caption{\label{fig:ftprof}Dynamic average treatment effects on the percentage of female employed in the professional occupation group. Event time is relative to the announcement of board gender diversity policies. All values are shown in percent.}
\end{figure}

We consider effects on the extensive margin, on full time employment. Figure \ref{fig:ftprof} shows the effect on the percentage of female among full time professional employees. Both quota and disclosure leads to a significant increase in the percentage of female professionals, reaching around 2\% to 3\% after five quarters. The effects are not significantly different between the two policies. 

\section{Conclusion}
In this paper, we propose a new synthetic control method to study effects of policies with staggered adoption. Our method overcomes limitations in existing methodologies, and gives asymptotically unbiased estimators of many interesting quantities and delivers asymptotically valid inference. There is potential for a wide application of this staggered synthetic control, since many policies are replicated in other regions at different timings. 

We apply the staggered synthetic control method to study how corporate board gender policies affect long term labor outcomes related to gender equality. By exploiting variation in the staggered announcement of board gender policies across European countries, we can estimate the dynamic treatment effect on gender and labor outcomes. 

Our paper sheds light on the extent to which the board gender policies enhance gender equality in the labor market. We find some evidence that these policies lead to an increase in the percentage of female professionals. This may suggest the higher female representation at the board level motivates other female employees to reach for the glass ceiling.  

While prior paper examining board gender policies focus on the short term firm level impact, we provide evidence on the long term societal impact, which is the main objective of these policies. This is feasible partly because of our proposed methodology using staggered synthetic control. Using staggered synthetic control has various advantages over existing methods to study this research question with aggregated country-level data and staggered adoption of policies. In particular, we mitigate concerns about the violation of parallel trend assumption in small sample setting, and we use all important units in forming the synthetic control, including treated units.


\appendix
\numberwithin{equation}{section} 

\section{Proofs of Results}
\renewenvironment{proof}{{\bfseries Proof of Theorem \ref{estimation_theorem}.}}{\qed}
\noindent\begin{proof}
	Using Equation \ref{estimator_formula}, we have 
\begin{align*}
\widehat{\tau}
=&\Bigg( \sum_{s=1}^S A_s' \widehat{M} A_s \Bigg)^{-1} \Bigg( \sum_{s=1}^S A_s'(I-\widehat{B})'((I-\widehat{B})\tau_s+Y_{T+s}(0)-\widehat{B}Y_{T+s}(0)-\widehat{a}) \Bigg) \\
=& \tau+V_T+\Bigg( \sum_{s=1}^S A_s' \widehat{M} A_s \Bigg)^{-1} \Bigg( \sum_{s=1}^S A_s'(I-\widehat{B})'((B-\widehat{B})Y_{T+s}(0)+(a-\widehat{a})) \Bigg)\\
=& \tau + V_T+ \Bigg( \sum_{s=1}^S A_s' {M} A_s +o_{p}(1) \Bigg)^{-1} \Bigg( \sum_{s=1}^S A_s'(I-{B}+o_{p}(1))'(o_{p}(1)+o_{p}(1)) \Bigg)\\
=& \tau + V_T + o_{p}(1).
\end{align*}
The first equality is by the definition of $\tau_s$. 
The third and last equations are by Assumption \ref{identification_assumption} and \ref{unbiased_assumption}. Therefore,
$$\widehat{\gamma}-(\gamma+LV_T)=L(\widehat{\tau}-\tau-V_T)=o_{p}(1).$$
In addition, $E[V_T]=0$ by Assumption \ref{unbiased_assumption}(a).
\end{proof}

\medskip

\renewenvironment{proof}{{\bfseries Proof of Theorem \ref{inference_theorem}.}}{\qed}
\noindent\begin{proof}
	We follow the proof of Theorem 2 in \citet{Andrews2006}. We use fours steps to show the theorem. 

\noindent \underline{Step 1} 
We first show $\widehat{P} \stackrel{d}{\rightarrow}  P_\infty$, where $P_\infty$ has the same distribution as $V_1(\theta_0)'C'CV_1(\theta_0)$. Using the result of Theorem \ref{estimation_theorem} and letting $L$ be the identity matrix, we have 
$$\widehat{\tau}-\tau=V_T+o_{p}(1) \stackrel{d}{\rightarrow}  V_1(\theta_0),$$
so under the null, $$C\widehat{\tau}-d=(C\widehat{\tau}-d)-(C\tau-d) \stackrel{d}{\rightarrow}  CV_1(\theta_0).$$
Applying the continuous mapping theorem, we have $\widehat{P} \stackrel{d}{\rightarrow}  P_\infty$. 

\medskip

\noindent \underline{{Step 2}} 
Let $F(x)$ and $q_{1-\alpha}$ be the distribution function and the $(1-\alpha)$-quantile of $P_\infty$, respectively. Next, we show $\widehat{F}(x)\stackrel{p}{\rightarrow} F(x)$ for all $x$ in a neighborhood of $q_{1-\alpha}$. 

Define $W_s=\Big( \sum_{r=1}^S A_r'MA_r \Big)^{-1}A_s'(I-B)' $
and its sample analog
$$\widehat{W}_s=\Bigg( \sum_{r=1}^S A_r'\widehat{M}A_r \Bigg)^{-1}A_s'(I-\widehat{B})'.$$
Let 
\begin{equation*}
\begin{aligned}
&L_{1,T}(\varepsilon)=\Big\lbrace \Vert (\widehat{\theta}-\theta_0)D_T\Vert_F\le \varepsilon, \max_{t=1,\dots,T}\Vert (\widehat{\theta}^{(t)}-\theta_0) D_T\Vert_F \le \varepsilon \Big\rbrace, \\
&L_{2,T}(c)=\Big\lbrace \max_{t\le T+S}\Vert D^{-1}_Tx_t\Vert\le c \Big\rbrace,\\
&L_{3,T}(\eta)=\Big\lbrace \forall r,s \text{ s.t. }1\le r\le S, 1\le s\le S, \Vert \widehat{W}_r'C'C\widehat{W}_s-{W}_r'C'C{W}_s\Vert_F<\eta \Big\rbrace.
\end{aligned}
\end{equation*}

By Assumption \ref{inference_assumption}(c), there exists a positive sequence $\{\varepsilon_T \}_{T\ge 1}$ such that $\varepsilon_T\rightarrow 0$ and $\Pr(L_{1,T}(\varepsilon_T))\rightarrow 1$. Let $c_T=1/\sqrt{\varepsilon_T}$. So we have $c_T\rightarrow \infty$ and $c_T\varepsilon_T\rightarrow 0$. By Assumption \ref{inference_assumption}(b), we must have $\Pr(L_{2,T}(c_T))\rightarrow 1$. By Assumption \ref{inference_assumption}(b), there exists a positive sequence $\{\eta_T \}_{T\ge 1}$ such that $\eta_T\rightarrow 0$ and $\Pr(L_{3,T}(\eta_T))\rightarrow 1$. Let $L_T=L_{1,T}(\varepsilon_T)\cap L_{2,T}(c_T)\cap L_{3,T}(\eta_T)$, then we have $\Pr(L_T)\rightarrow 1$ and $\Pr(L_T^c)\rightarrow 0$. 

Suppose $L_T$ holds. Then, for some $\theta=\widehat{\theta}$ or $\theta=\widehat{\theta}^{(t)}$ and for some $t=1,\dots, T$, we have 
\begin{equation}
\label{P_hat diff}
|\widehat{P}_t(\theta)-P_t(\theta_0)|\le |\widehat{P}_t(\theta)-P_t(\theta)|+|{P}_t(\theta)-P_t(\theta_0)|.
\end{equation}
Note that 
\begin{align}
\label{P_hat diff_1}
&|\widehat{P}_t(\theta)-P_t(\theta)|\notag\\
=& \Bigg| \sum_{r=1}^S\sum_{s=1}^S (Y_{t+r}-\theta x_{t+r})'(\widehat{W}_r'C'C\widehat{W}_s-{W}_r'C'C{W}_s)(Y_{t+s}-\theta x_{t+s}) \Bigg| \notag\\
\le& \sum_{r=1}^S\sum_{s=1}^S  \Vert Y_{t+r}-\theta x_{t+r}\Vert \Vert \widehat{W}_r'C'C\widehat{W}_s-{W}_r'C'C{W}_s\Vert_F \Vert Y_{t+s}-\theta x_{t+s}\Vert\notag\\
\le& \sum_{r=1}^S\sum_{s=1}^S  \Vert u_{t+r}+(\theta_0-\theta)x_{t+r}\Vert \cdot \eta_T\cdot \Vert u_{t+s}+(\theta_0-\theta)x_{t+s}\Vert\notag\\
\le& \sum_{r=1}^S\sum_{s=1}^S (\Vert u_{t+r}\Vert+\Vert (\theta_0-\theta)D_TD_T^{-1}x_{t+r}\Vert)\eta_T(\Vert u_{t+s}\Vert+\Vert (\theta_0-\theta)D_TD_T^{-1}x_{t+s}\Vert)\notag\\
\le&  \sum_{r=1}^S\sum_{s=1}^S(\Vert u_{t+r}\Vert +\Vert (\theta_0-\theta)D_T\Vert_F \Vert D_T^{-1}x_{t+r}\Vert)\eta_T(\Vert u_{t+s}\Vert +\Vert (\theta_0-\theta)D_T\Vert_F \Vert D_T^{-1}x_{t+s}\Vert)\notag\\
\le& \sum_{r=1}^S\sum_{s=1}^S(\Vert u_{t+r}\Vert +\varepsilon_Tc_T)(\Vert u_{t+s}\Vert +\varepsilon_Tc_T)\eta_T
\end{align} 
and
\begin{align}
\label{P_hat diff_2}
& |{P}_t(\theta)-P_t(\theta_0)|\notag \\
\le& \sum_{r=1}^S\sum_{s=1}^S  |(Y_{t+r}-\theta x_{t+r})'{W}_r'C'C{W}_s(Y_{t+s}-\theta x_{t+s})-(Y_{t+r}-\theta x_{t+r})'{W}_r'C'C{W}_s(Y_{t+s}-\theta_0 x_{t+s})|\notag\\
&+	|(Y_{t+r}-\theta x_{t+r})'{W}_r'C'C{W}_s(Y_{t+s}-\theta_0 x_{t+s})-(Y_{t+r}-\theta x_{t+r})'{W}_r'C'C{W}_s(Y_{t+s}-\theta x_{t+s})|\notag\\
=& \sum_{r=1}^S\sum_{s=1}^S |(u_{t+r}+(\theta_0-\theta) x_{t+r})'{W}_r'C'C{W}_s(\theta_0-\theta) x_{t+s}|+|((\theta_0-\theta)x_{t+r})'{W}_r'C'C{W}_su_{t+s}|\notag\\
\le &  \sum_{r=1}^S\sum_{s=1}^S  (\Vert u_{t+r}\Vert+\Vert u_{t+s}\Vert +\varepsilon_Tc_T) \Vert {W}_r'C'C{W}_s\Vert_F \varepsilon_Tc_t.
\end{align}
Combining \eqref{P_hat diff}, \eqref{P_hat diff_1}, and \eqref{P_hat diff_2}, we have 
\begin{equation*}
|\widehat{P}_t(\theta)-P_t(\theta_0)|\le g(\varepsilon_T,c_T,\eta_T),
\end{equation*}
where
\begin{align*}
&g_t(\varepsilon_T,c_T,\eta_T)\\
=  & \sum_{r=1}^S\sum_{s=1}^S(\Vert u_{t+r}\Vert +\varepsilon_Tc_T)(\Vert u_{t+s}\Vert +\varepsilon_Tc_T)\eta_T+(\Vert u_{t+r}\Vert+\Vert u_{t+s}\Vert +\varepsilon_Tc_T) \Vert {W}_r'C'C{W}_s\Vert_F \varepsilon_Tc_t.
\end{align*}
By Assumption \ref{unbiased_assumption}(a), $g_t(\varepsilon_T,c_T,\eta_T)$ is identically distributed across $t$ for a fixed $T$.

Let $k:\mathbb{R}\rightarrow\mathbb{R}$ be a monotonically decreasing and everywhere differentiable function that has bounded derivative and satisfies $k(x)=1$ for $x\le 0$, $k(x)\in [0,1]$ for $x\in (0,1)$, and $k(x)=0$ for $x\ge 1$. For example, let $k(x)=\cos(\pi x)/2+1/2$ for $x\in (0,1)$. Given some sequence $\{p_t \}_{t=1}^T$, a smoothed distribution function is defined by
\begin{equation*}
\tilde{F}(x,\{p_t\},h_T)=\frac{1}{T}\sum_{t=1}^{T}k\bigg( \frac{p_t-x}{h_T} \bigg), 
\end{equation*}
for some sequence of positive constants $\{h_T \}$ such that $h_T\rightarrow 0$, $c_T\varepsilon_T/h_T\rightarrow 0$, and 
$\eta_T/h_T\rightarrow 0$. For example, we can let $h_T=\max \{\sqrt{\varepsilon_T c_T},\sqrt{\eta_T} \}$. 

Define ${F}_T(x)=\frac{1}{T}\sum_{t=1}^{T-S}\mathbbm{1}\{P_t(\theta_0)\le x \}$. We write
\begin{equation*}
|\widehat{F} (x)-F (x)|\le \sum_{i=1}^{4}D_{i,T},
\end{equation*}	
for 
\begin{align*} 
D_{1,T}&= |\widehat{F} (x)-\tilde{F}(x,\{\widehat{P}_t \},h_T)|,\notag\\ 
D_{2,T}&= |\tilde{F}(x,\{\widehat{P}_t \},h_T)-\tilde{F}(x,\{ P_t(\theta_0) \},h_T)|,\notag\\ 
D_{3,T}&= |\tilde{F}(x,\{P_t(\theta_0) \},h_T)-{F}_T(x)|,\text{ and}\notag\\ 
D_{4,T}&= |{F}_T(x)-{F}(x)|. 
\end{align*}
We want to show that all four terms vanish. First note that 
\begin{equation*}
D_{1,T}\le \frac{1}{T}\sum_{t=1}^{T}\mathbbm{1}\Bigg\lbrace \frac{\widehat{P}_t(\widehat{\theta}^{(t)})-x}{h_T}\in (0,1) \Bigg\rbrace .
\end{equation*} 
Thus, for any $\delta>0$, 
\begin{align}
\label{bound D1T} 
\Pr(D_{1,T}>\delta)&\le \Pr(\{D_{1,T}>\delta \}\cap L_T)+\Pr(L_T^c)\notag\\
&\le \Pr \Bigg( \frac{1}{T}\sum_{t=1}^{T}\mathbbm{1}\big\lbrace {P}_t({\theta}_0)-x\in (-g_t(\varepsilon_T,c_T,\eta_T),h_T+g_t(\varepsilon_T,c_T,\eta_T) \Bigg\rbrace>\delta \big) +o(1)\notag\\
&\le \frac{E\mathbbm{1}\big\lbrace P_t(\theta_0)-x\in (-g_t(\varepsilon_T,c_T,\eta_T),h_T+g_t(\varepsilon_T,c_T,\eta_T) \big\rbrace}{\delta}+o(1),
\end{align}
where the last inequality is by Markov's inequality. Recall $\Pr(P_1(\theta_0)\ne x)=1$ and $g_t(\varepsilon_T,c_T,\eta_T)\rightarrow 0$ almost surely, so $\mathbbm{1}\{P_t(\theta_0)-x\in \{-g_t(\varepsilon_T,c_T,\eta_T),h_T+g_t(\varepsilon_T,c_T,\eta_T)  \}\rightarrow 0$ almost surely. By the dominated convergence theorem, \eqref{bound D1T} implies $\Pr(D_{1,T}>\delta)\le o(1)$ and thus $D_{1,T}=o_{p}(1)$. 
For $D_{2,T}$, we have
$$D_{2,T}= \Bigg| \frac{1}{T} \sum_{t=1}^{T} k'\Bigg( \frac{\tilde{P}_t-x}{h_T} \Bigg) \frac{\widehat{P}_t(\widehat{\theta}^{(t)})-P_t(\theta_0)}{h_T} \Bigg| \le \frac{\bar{k}}{T}\sum_{t=1}^{T}\frac{g_t(\varepsilon_T,c_T,\eta_T)}{h_T}.$$
The equality is by the mean value theorem and we have $\tilde{P}_t$ lies between $\widehat{P}_t(\widehat{\theta}^{(t)})$ and $P_t(\theta_0)$. In the inequality, $\bar{k}$ is a bound for the derivative of $k$. Also, note
\begin{align*}
&E\bigg[ \frac{g_t(\varepsilon_T,c_T,\eta_T)}{h_T}\bigg]\\
=&\sum_{r=1}^S\sum_{s=1}^S(\Vert u_{t+r}\Vert +\varepsilon_Tc_T)(\Vert u_{t+s}\Vert +\varepsilon_Tc_T)\eta_T+(\Vert u_{t+r}\Vert+\Vert u_{t+s}\Vert +\varepsilon_Tc_T) \Vert {W}_r'C'C{W}_s\Vert_F \varepsilon_Tc_t\\
=&\sum_{r=1}^S\sum_{s=1}^S\Bigg( \frac{E[\Vert u_{t+r}\Vert \Vert u_{t+s}\Vert ]\eta_T}{h_T}+\frac{2E[\Vert u_{t+r}\Vert ]\varepsilon_T c_T \eta_T}{h_T}+\frac{\varepsilon_T^2 c_T^2\eta_T}{h_T}\Bigg.\\
&\Bigg. +\frac{2E[\Vert u_{t+r}\Vert] \Vert W_r'C'CW_s\Vert_F \varepsilon_T c_T}{h_T}+\frac{\varepsilon_T^2 c_T^2 \Vert W_r'C'CW_s\Vert_F}{h_T}\Bigg) \\
=&o(1).
\end{align*}
Therefore, 
\begin{align*}
\Pr(D_{2,T}>\delta)&\le \Pr (\{D_{2,T}>\delta \}\cap L_T)+\Pr (L_T^c)\notag\\
&\le \Pr \Bigg( \frac{\bar{k}}{T}\sum_{t=1}^{T}\frac{g_t(\varepsilon_T,c_T,\eta_T)}{h_T}>\delta \Bigg) +o(1)\notag\\
&\le \bar{k} \frac{Eg_t(\varepsilon_T,c_T,\eta_T)}{\delta h_T}\notag\\
&\rightarrow 0. 
\end{align*}
The third inequality is by Markov's inequality. This shows $D_{2,T}=o_{p}(1)$. 

$D_{3,T}$ is similar to the $D_{1,T}$ case. Finally, by stationary and ergodicity of $u_{t}$, we have $D_{4,T}=o_{p}(1)$. This implies $\widehat{F}(x)\stackrel{p}{\rightarrow} F(x)$. 

\medskip

\noindent \underline{Step 3} 
Now we show $\widehat{q} _{1-\alpha}\stackrel{p}{\rightarrow} q _{1-\alpha} $. Pick any small $\varepsilon$ such that $\widehat{F} (x)\stackrel{p}{\rightarrow}F (x)$ for $x\in (q _{1-\alpha}-\varepsilon,q _{1-\alpha}+\varepsilon)$. Note 
\begin{align*}
\Pr (\widehat{q} _{1-\alpha}>q _{1-\alpha}+\varepsilon)
&\le \Pr (\widehat{F} ({q} _{1-\alpha}+\varepsilon)<1-\alpha)\notag\\
&= \Pr (\widehat{F} ({q} _{1-\alpha}+\varepsilon)-{F} ({q} _{1-\alpha}+\varepsilon)<(1-\alpha)-{F} ({q} _{1-\alpha}+\varepsilon))\notag\\
&\rightarrow 0.
\end{align*}
The inequality is by definition of $\widehat{q} _{1-\alpha}$. The convergence is by Assumption \ref{inference_assumption}(d) and Step 2. Similarly, 
\begin{align*}
\Pr (\widehat{q} _{1-\alpha}<q _{1-\alpha}-\varepsilon)
&\le \Pr (\widehat{F} ({q} _{1-\alpha}-\varepsilon)\ge 1-\alpha)\notag\\
&= \Pr (\widehat{F} ({q} _{1-\alpha}-\varepsilon)-{F} ({q} _{1-\alpha}-\varepsilon)\ge (1-\alpha)-{F} ({q} _{1-\alpha}-\varepsilon))\notag\\
&\rightarrow 0.
\end{align*}
Thus, $\Pr(|\widehat{q} _{1-\alpha}-q _{1-\alpha}|> \varepsilon)\rightarrow 0$. 

\medskip

\noindent \underline{Step 4}
Finally, we show $\Pr(\widehat{P}>\widehat{q} _{1-\alpha})\rightarrow \alpha$. Under null, we have
\begin{align*}
\Pr(\widehat{P}>\widehat{q} _{1-\alpha})&=1-\Pr (\widehat{P}\le \widehat{q} _{1-\alpha})\notag\\
&=1-\Pr (\widehat{P}+( {q} _{1-\alpha}- \widehat{q} _{1-\alpha})\le {q} _{1-\alpha})\notag\\
&\rightarrow \alpha,
\end{align*}
where the convergence is by combining Step 1 and 3. This concludes our proof.
\end{proof}

\

\renewenvironment{proof}{{\bfseries Proof of Corollary \ref{inference_att}.}}{\qed}
\noindent\begin{proof}
This corollary is a direct application of Theorem \ref{estimation_theorem} where $L=l_s'$.
\end{proof}

\section{Primitive Assumptions}
\label{low_level_section}
In this section we provide a set of low-level conditions under which Assumption \ref{unbiased_assumption} and \ref{inference_assumption} hold. Following \citet{Ferman2021} and \citet{Cao2019}, we consider a factor model such that for $i=1,\dots,N$ and $t=1,\dots,T+S$,  
\begin{equation}
\label{foctor model form}
y_{i,t}(0)=\lambda_i'f_t+\varepsilon_{i,t},
\end{equation}
where $f_t$ is finite-dimensional common factors. For notation simplicity, we write $Y_t(0) = (y_{1,t}(0),\dots,y_{N,t}(0))'$, $Y_t = (y_{1,t},\dots,y_{N,t})'$, and $\varepsilon_t=(\varepsilon_{1,t},\dots,\varepsilon_{N,t})'$. 

\begin{condition}[model with stationary common factors]
\label{cond st}
	Assume $\{(f_t,\varepsilon_t)\}_{t\ge 1}$ is stationary, ergodic for the first and second moments, and has finite $(2+\delta)$-moment for some $\delta>0$. Assume $cov[Y_t(0)]=\Omega_y$ is  positive definite.
\end{condition}

\begin{condition}[model with cointegrated $\mathcal{I}(1)$ common factors] 
\label{cond co}
Rewrite Equation \eqref{foctor model form} as 
	\begin{equation*}
	y_{i,t}(0)=	(\lambda_i^1)'f_t^1+	(\lambda_i^0)'f_t^0+\varepsilon_{i,t}.
	\end{equation*}
Assume $\{(f_t^0,\epsilon_t) \}_{t\ge 1}$ is stationary, ergodic for the first and second moments, and has finite $4$-th moment. Without loss of generality, $E[\varepsilon_{i,t}]=0$. Assume $\{f_t^1\}_{t\ge 1}$ is $\mathcal{I}(1)$. Further assume for each $i$, $y_{i,t}(0)$ is such that weak convergence holds for $T^{-1/2}y_{i,[rT]}(0)\Rightarrow \nu_i(r)$, where $\Rightarrow$ is weak convergence and process $\nu_i(r)$ is defined on $[0,1]$ and has bounded continuous sample path almost surely.  For each $i$, let $W^{(i)}=\{(w_1,\dots,w_N)\in \mathbb{R}_+^N:w_i=0,\sum_{j\ne i}w_j=1 \}$. Assume for each $i$, there exists $w^{(i)}\in W^{(i)}$ such that $\lambda_i^1=\sum_{j=1}^Nw^{(i)}_j\lambda_j^1$. That is, $(w^{(i)}-e_i)$ is a cointegrating vector for $Y_t(0)$, where $e_i$ is a unit vector with $i$-th entry being one and zeros everywhere else.
\end{condition}

The following lemma shows that under the factor model, either stationarity or co-integration implies the high-level assumptions in the paper. 
\begin{lemma}
	\label{lemma_low_level_assumptions}
	Suppose the distribution function of $P_1(\theta_0)$ is continuous and increasing at its $(1-\alpha)$-quantile. Then, either Condition \ref{cond st} or Condition \ref{cond co} implies Assumption \ref{unbiased_assumption} and \ref{inference_assumption}.
\end{lemma}

\renewenvironment{proof}{{\bfseries Proof.}}{\qed}
\noindent\begin{proof}
This lemma is implied by Lemma 1 and 3 of \citet{Cao2019}. The main idea is to use the projection of the least-square estimator onto a constraint set. 
\end{proof}
\FloatBarrier
\phantomsection
\addcontentsline{toc}{section}{References}
\bibliography{references}{} 
\bibliographystyle{aer}

\end{document}